# Physical Activity Recognition by Utilising Smartphone Sensor Signals

Abdulrahman Alruban[1, 2 ✉], Hind Alobaidi[1, 3], Nathan Clarke[1, 4], Fudong Li[1, 5]

[1]*Centre for Security, Communications and Network Research, Plymouth University*
*Plymouth, UK*
[2]*Computer Sciences and Information Technology College, Majmaah University*
*Majmaah, Saudi Arabia*
[3]*College of Education for Pure Science, University of Baghdad*
*Baghdad, Iraq*
[4]*Security Research Institute, Edith Cowan University*
*Perth, Western Australia*
[5]*School of Computing, University of Portsmouth*
*Portsmouth, UK*
*{abdulrahman.alruban, hind.al-obaidi, n.clarke, fudong.li}@plymouth.ac.uk*

Abstract: Human physical motion activity identification has many potential applications in various fields, such as medical diagnosis, military sensing, sports analysis, and human-computer security interaction. With the recent advances in smartphones and wearable technologies, it has become common for such devices to have embedded motion sensors that are able to sense even small body movements. This study collected human activity data from 60 participants across two different days for a total of six activities recorded by gyroscope and accelerometer sensors in a modern smartphone. The paper investigates to what extent different activities can be identified by utilising machine learning algorithms using approaches such as majority algorithmic voting. More analyses are also provided that reveal which time and frequency domain-based features were best able to identify individuals' motion activity types. Overall, the proposed approach achieved a classification accuracy of 98% in identifying four different activities: walking, walking upstairs, walking downstairs, and sitting (on a chair) while the subject is calm and doing a typical desk-based activity.

## 1. INTRODUCTION

Human physical activity identification has gained considerable amount of attention due to the prevalent use of smartphone devices and motion sensing technology advancement that facilitates the monitoring of human activities by small portable devices. The majority of modern smartphones have a number of built-in sensors (e.g., GPS, accelerometers, magnetometers, gyroscopes, barometers, temperature and humidity sensors) that can be utilised to record a variety of individuals' activity signals. This has enabled research in activity-based computing to become a cornerstone of many real-life applications in health care, the military, navigation, localisation, biometrics, sport analytics and security (He and Li, 2013; Mitchell, Monaghan and O'Connor, 2013; Bayat, Pomplun and Tran, 2014; Al-Naffakh *et al.*, 2016; Ronao and Cho, 2016). Researchers in the field of human activity identification have utilised a number of techniques to enhance the accuracy of activity type recognition, mostly based on acceleration and angular velocity signals using accelerometer and gyroscope sensors embedded in mobile devices. The sensors generate tri-axial linear signals which can be processed and segmented into less noisy features that provide a latent pattern that captures the context of the motion activity type. Prior research has focused mainly on the performance of the approaches developed in solving a particular problem, such as activity identification, which has meant that there has been little focus on interpreting how the identification decision was made in the case of machine learning modelling (Lara and Labrador, 2013; Jiang and Yin, 2015; Ha and Choi, 2016). This includes investigating which feature contributed the most to the identification (prediction) process. The majority of human activity recognition public datasets upon which much of the literature has been built also have a limited number of participants and samples (Altun, Barshan and Tunçel, 2010; Anguita *et al.*, 2012; Reyes-Ortiz *et al.*, 2014). This presents challenges to understanding whether the captured activity signals of an individual vary over time (e.g., across days), as, typically, most of these datasets were collected on the same day.

Therefore, this study investigates the effect of using a feature ranking approach prior to the activity identification process by utilising random forest classification (Palczewska *et al.*, 2014), as this algorithm analyses which independent variable(s) contributed most during the training phase of a learning algorithm. This is undertaken by examining a dataset of 60 participants, which was collected for this study over two days. Finally, the proposed approach is evaluated by building a predictable model that is able to categorise a given individual's activity signals into predefined classes (i.e., normal walk, fast walk, walk with bag, walk upstairs, walk downstairs, and sitting). The modelling utilises three supervised machine learning classification algorithms: eXtreme Gradient Boosting (XGB), a feedforward neural network (NN) and a support vector machine (SVM).

The rest of the paper is organised as follows: Section 2 highlights related work in the area of human activity identification using mainly smartphone sensors. Section 3 explains the data collection and experimental methodology. Section 4 presents the experimental results of the different tests undertaken to evaluate the proposed approach. Section 5 discusses the findings and possible future work. The paper concludes in Section 6.

## 2. BACKGROUND AND RELATED WORK

Human activity recognition is a wide research field and studies in this area vary in a number of aspects. For instance, some studies can be categorised based on the way the data was collected, such as those using wearable sensors (e.g., smartwatch-mounted body devices) or smartphone devices, while other studies use video observation to record individuals' activity signals. With respect to devices such as smartphones, a key advantage is that the sensors are embedded, and no additional hardware is needed; only the software needs to be developed to start collecting activity motion signals. Therefore, much of the research has employed smartphones to record various types of individual activities in a user-friendly, unobtrusive, and periodic manner (Kwon, Kang and Bae, 2014; Capela *et al.*, 2016; Shoaib *et al.*, 2016). Most of the studies that have utilised smartphone-embedded sensors place the device either in a pouch or inside a trouser pocket (Ganti, Srinivasan and Gacic, 2010; Bieber *et al.*, 2011; Hamm *et al.*, 2013; Antos, Albert and Kording, 2014; Bahle *et al.*, 2014). In terms of activity recognition performance, in San-Segundo, Blunck, Moreno-Pimentel, Stisen, and Gil-Martín, 2018 study, the authors conducted a comprehensive evaluation of smartphone- and smartwatch-based human activity recognition, and found that smartphones mostly outperformed smartwatches in recognising activity type. This was due to the greater noise in the recordings from smartwatch sensors. Typically, both devices record activity signal data using the tri-axis signals of accelerometer and gyroscope sensors at a sampling rate ranging from 20 to 50 signals per second.

A number of approaches are used for data pre-processing and feature extraction, including cycle-based, segment-based and deep learning algorithms. In a cycle-based approach, the captured activity data are supposed to be a periodic signal in which each cycle begins once a foot touches the ground and finishes when the same foot touches the ground for the second time (i.e., two steps for a human) (Derawi and Bours, 2013). In a segment-based method, the signals are divided into fixed time-length windows (e.g., 10 seconds). Some gait activities are periodic, as each time segment is reasonably assumed to contain similar signal features, while some activity streams, such as standing and sitting, do not necessarily generate cycle-like patterns. In addition, the segmenting of the signals based on a time sequence requires fewer computational operations than the cycle-based method does.

In contrast with the cycle- and segment-based approaches, some researchers have utilised deep learning to meet the challenges of the feature extraction process. With the recent advances in deep learning algorithms, the use of convolutional neural network (CNN) learning algorithms to extract a latent pattern from raw data has become common practice (Jiang and Yin, 2015; Ronao and Cho, 2016). Typically, deep learning approaches require less effort in feature extraction and engineering in comparison with cycle- and segment-based approaches. However, a challenging aspect in deep learning-based models is that it is hard to explain and interpret how decisions are made (Weld and Bansal, 2018). Knowing what drives decisions in models (i.e., the features on which the model relies) is an important element in some activity recognition applications, such as health care-related research.

In (Kwapisz, Weiss and Moore, 2010), the study used a neural network to model human activity and achieved high accuracy in identifying the correct class to which the activity signals belonged. However, the limited number of population samples (i.e., 5-30) opens the possibility that the learned algorithm is overfitted and has memorised the training samples.

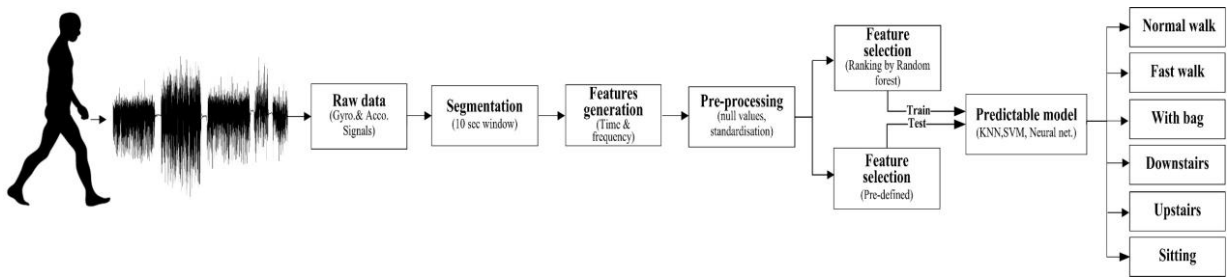

Figure 1: Data process pipeline.

Other studies, as shown in Table 1 (Ganti, Srinivasan and Gacic, 2010; Anguita *et al.*, 2012; Nakano, 2017; Bhanu Jyothi and Hima Bindu, 2018; Ogbuabor and La, 2018), have used a sliding window approach with an overlap of 50% in segmenting the raw activity signals. This could, however, lead to an overlap in the subsampling between the training and testing sets, which means that unless the splitting of the two sets occurs before the segmenting of the raw data, the data are only partially seen by the learning algorithm in both the training and testing sets. In terms of the correct classification rate, it can be seen that SVM, neural network and CNN achieve the highest performance among the techniques shown.

In this study, a segment-based approach is used to extract features from raw sensor signal data with a sliding window of 10 seconds with no overlap. The extracted features are used to compute various statistical features, such as the mean, median, maximum and minimum of a given sensor axis within a specific segment window (as explained in detail in section 3). By handcrafting these features, it is possible to understand which of the features contributed most effectively to discriminating individuals' activities (as presented in Section 4). In comparison with existing studies in which the data were gathered from smartphones, as presented in Table 1, most of these studies have fewer participants, (i.e., 30 or fewer) and the data were all captured during the same day. In this study, the data were collected between two days for everyone within the sample set because the probability that users' activity patterns change is higher for data collected across days than it is for data gathered on the same day.

Table 1: Comparison of prior studies in activity recognition using smartphone sensors.

| Study | Approach | Performance (CCR%) | Population | Activity Type |
|---|---|---|---|---|
| (Kwapisz, Weiss and Moore, 2010) | NN | 100 | 5 | Standing, sitting, walking, jogging, downstairs, upstairs |
| (Anguita *et al.*, 2012) | SVM | 89 | 30 | Standing, sitting, walking, lying down, downstairs, upstairs |
| (Ganti, Srinivasan and Gacic, 2010) | SVM | 96 | | |
| (Nakano, 2017) | CNN | 90 | | |
| (Bhanu Jyothi and Hima Bindu, 2018) | RF PCA | 94 89 | | |
| (Ogbuabor and La, 2018) | MLP | 95 | | |
| (Jiang and Yin, 2015) | CNN | 99 | 10 | Standing, sitting, walking, jogging, running, biking, downstairs, upstairs |
| (Heng, Wang and Wang, 2016) | SVM | 85 | 5 | Standing, walking, running, upstairs, downstairs |
| (Saha *et al.*, 2018) | Ensemble | 94 | 10 | Sitting on a chair, sitting on the floor, lying right, lying left, slow walk, brisk walk |

Legend: CCR: correct classification rate; ML: machine learning; PCA: principal component analysis; MLP: Multi-layer perceptron; RF: random forest.

## 3. METHODOLOGY

This study follows the data pipeline flow presented in Figure 1. The overall process starts by capturing the raw activity signals from smartphone sensors, followed by segmenting the data with time windows of 10 seconds. Once the data are segmented, they are processed to extract statistical features. This is followed by standardising feature space values and ranking those features using random forest algorithms. After that, the activity samples are fitted into the learning algorithm to train the predictable model. The process is explained in more detail in the following subsections.

### 3.1 Data Collection Sensors and Device

The developed approach utilises embedded smartphone motion sensors: the gyroscope and the accelerometer. A gyroscope is used to maintain a reference direction in the motion systems by sensing the degree of orientation in the x, y, and z directions of the smartphone. The axis signal is affected by the direction of the device orientation. Also, the accelerometer sensor measures the acceleration in metres per second squared (m/s$^2$) in the x, y, and z directions of the smartphone. Figure 3 show the orientation of the positive and negative x, y, and z-axes for a typical smartphone device using the gyroscope and accelerometer sensors respectively. An Android application called AndroSensor was used to record the sensor data as it supports most of the sensors that an Android device can offer (F, no date). A Samsung Galaxy S6 smartphone was carried by each individual to record the sensor data generated by different human physical activities. Each user was asked to place the smartphone in a belt pouch, as presented in Figure 4. The generated data were continuously collected at a rate of 30-32 Hz for the x, y, and z-axes of both the accelerometer and gyroscope sensors.

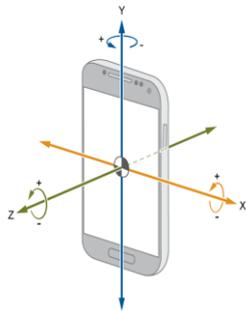
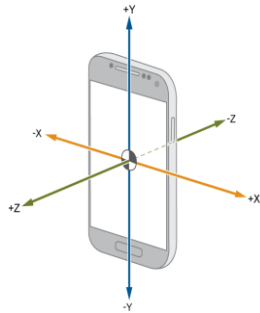

Figure 2: Orientation of the axes relative to a typical smartphone device using a gyroscope sensor.

Figure 3: Orientation axes relative to a typical smartphone device using an accelerometer sensor.

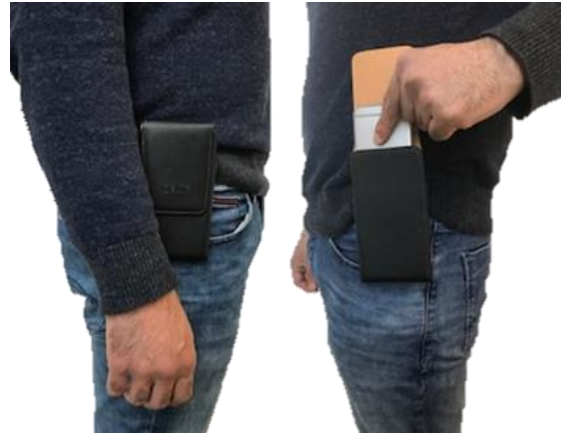

Figure 4: Smartphone device located inside a pouch.

### 3.2 Data Collection Scenarios

During the data collection process, each individual was asked to walk normally, fast, and normally with a bag on a predefined route (along a flat corridor) for a period of 3 minutes for each activity. For more realistic scenarios, the participant had to stop to open a door and walk back and forth along the corridor a number of times. This was followed by walking downstairs for three levels and upstairs for the same three levels, which resulted in a total number of 126 steps (63 for each direction). Between each activity, the participant was asked to stop for 15 to 20 seconds to rest as well as to allow the later manual separation of the generated signals into their corresponding activities. Ten sessions of user activities were collected per user: five sessions were from one day, and the other five sessions were collected one week later from the same participant, in addition to a sitting activity for 19 of the participants. The users were permitted to wear different footwear and clothing for the second day of data collection. In total, 60 users participated in the data collection exercise; 35 participants were male and 25 were female, and they were aged between 18 and 56 years old.

Upon completion of the data collection phase, users' activities were divided into six datasets aligned to each activity (i.e. normal walk, fast walk, walk with a bag, downstairs walking, upstairs walking, and sitting). Then the tri-axial raw accelerometer and gyroscope signals were segmented into 10-second segments using a sliding window approach with no overlapping to compute the feature set that is explained in the next subsection.
Table 2 shows the collected dataset information.

Table 2: Dataset information.

| Activity type | #User | #Raw signal samples per user (~32\sec) | Processed #samples per user | #Seconds per user | Total #samples |
|---|---|---|---|---|---|
| Normal | 60 | 7,168 | 28 | 280 | 1,680 |
| Fast | 60 | 7,424 | 29 | 290 | 1,740 |
| W/bag | 60 | 6,912 | 27 | 270 | 1,620 |
| Downstairs | 60 | 1,792 | 7 | 70 | 436 |
| Upstairs | 60 | 1,536 | 6 | 60 | 410 |
| Sitting | 19 | 4,096 | 52 | 160 | 997 |

## 3.3 Feature Extraction

The raw signal data generated by the gyroscope and accelerometer were processed by computing the time and frequency domain features as this is a standard approach to generating a feature vector. These features were extracted from the users' data segments. The time domain features were calculated directly from the raw data samples, while a Fourier transform was applied to the raw signals across the three sensor axes before computing the frequency domain-based features set. This process generated 304 unique features from the two domains, as listed in Table 3.

Table 3: Generated features.

| Feature domain | Feature type (count) |
|---|---|
| Time and frequency | Mean (3), standard deviation (3), median (3), variance (3), covariance (3), zero crossing rate minimum, interquartile range, average absolute, difference (3), root mean square (3), skewness (3), kurtosis (3), percentile 25 (3), percentile 50 (3), percentile 75 (3), maximum (3), minimum (3), correlation coefficients (3), average resultant acceleration (1) |
| Time only | Difference (3), maximum value (4), minimum value (4), binned distribution (3), maximum peaks (3), minimum peaks (3), peak occurrence (3), time between peaks (3), interquartile range (3) |
| Frequency only | Entropy (3), energy (3) |

## 3.4 Modelling

Data modelling aims to build a predictable model able to classify a given individual's activity signals into the class to which it belongs, based on the features extracted from the raw sensor data (in this case, normal walk, fast walk, walk with bag, downstairs, upstairs, and sitting). The following steps were undertaken before fitting the samples into the selected machine learning algorithms.

### 3.4.1 Data Pre-Processing

Two approaches (i.e., normalisation and standardisation) were examined for transforming data. The dataset was normalised by scaling the input vectors individually to the unit norm (vector length). The other transformation approach was to standardise the features by removing the mean and scaling to the unit variance. The latter approach (standardisation) emerged as better than the former (normalisation) in discriminating the activity samples for the tested dataset.

### 3.4.2 Feature Selection

In order to reduce the feature vector dimensions, only those ranked as being of higher importance in contributing most effectively to discriminating individuals' activities by the random forest algorithm were included in training the predictable model. The variable importance measure of the random forest calculates how significantly a given feature is biased towards correlated predictor variables (Strobl *et al.*, 2008). Feature importance analysis using random forest reduced the feature vector from 304 to 195 features in the final model based on the training set data. Reducing the feature space dimensionality not only improves overall model performance, but also lowers the probability of the algorithm being overfitted to the training data.

### 3.4.3 Train and Test Split Ratio

The cross-validation (CV) approach was used to train and validate the base model as non-stratified fashion. Using CV tends to decrease the probability of overfitting. The dataset was split into five consecutive folds without shuffling. Each fold was then used once as a validation while the remaining four folds formed the training set.

### 3.4.4 Classification Algorithms

Three supervised machine learning classification algorithms were examined using: NN, SVM and XGB. The XGN parameters are (n_estimators:500,

max_depth:3, min_samples_leaf=4, max_features: 0.2). The SVM parameters are (C=1.6, kernel:'rbf'). The NN algorithm was tweaked by hyper-parameter tuning, using a grid-search approach as shows in table 4.

Table 4: summary of neural network tuned parameters.

| Parameter | Value |
| --- | --- |
| #of epoch | 500 |
| #of hidden layers | 1 |
| #of hidden nodes | 130 |
| Dropout rate | 0.6 |
| Hidden activation function | Relu |
| Output activation function | Softmax |
| Kernel initialiser | Uniform |
| Loss function | Categorical cross entropy |
| Optimiser | Stochastic gradient descent (SGD) |

## 4. EXPERIMENTAL ANALYSIS

As the feature vector contains 304 features, dimensionality reduction helped in improving the overall model performance. Therefore, the random forest algorithm was used to rank the feature sets based on their contribution to the decision being made in predicting the target variable (activity) using the algorithm. The features space was fitted into the algorithm and performed a conventional multi-class classification task. Once the model was trained, querying the features importance variable resulted in a list of all the independent variables and their ranks to measure how significant the features are in discriminating the target classes (human physical activities in the context of this study). Figure 5illustrates the top 10 ranked features of all those examined using the random forest algorithm.

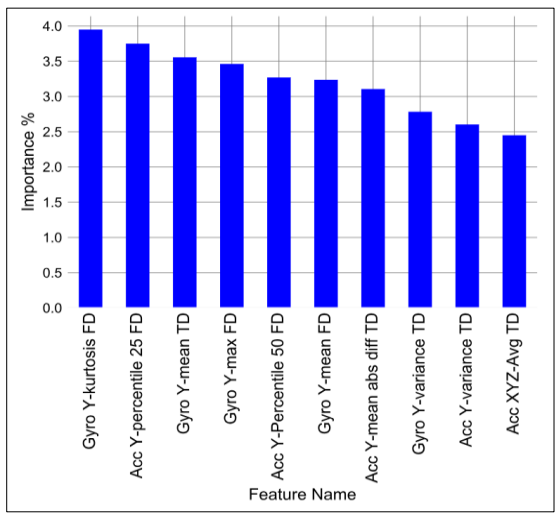

Figure 5: Top 10 features ranked using the random forest.

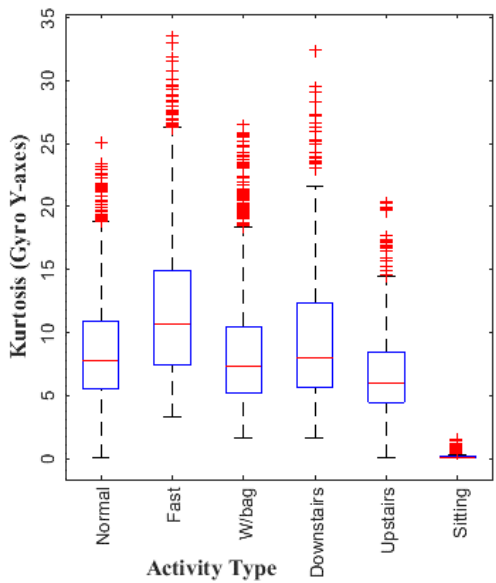

Figure 6: Kurtosis of the gyroscope y-axis.

Figure 6 illustrates the top-ranked feature, 'Kurtosis', which is a measure of the shape for the values in a particular segment. The plot depicts the Kurtosis data of the six activities through their quartiles. It is apparent from this descriptive statistic that there is clear variability across the activities examined for this feature. Although normal walk and walk with bag are two different activities, they are, by their nature, very similar in terms of pace and type of body movement. This is clearly seen in Figure 6, as the median and first and third quartiles are almost equal for this feature as computed by the random forest algorithm. When examining the confusion matrix for the predictable model (later in this section), most of the false positive samples are also between these two activities, which supports the point being made here.

In contrast, Figure 7 presents the lowest-ranked feature, which corresponds to the binned distribution of the minimum and maximum accelerations of the z-axes in the segments. Almost all the activity values of this feature are identical, except for sitting. This descriptive analysis visually validates the output of the algorithmic feature ranking approach as the top-ranked features have more variability than those ranked lower.

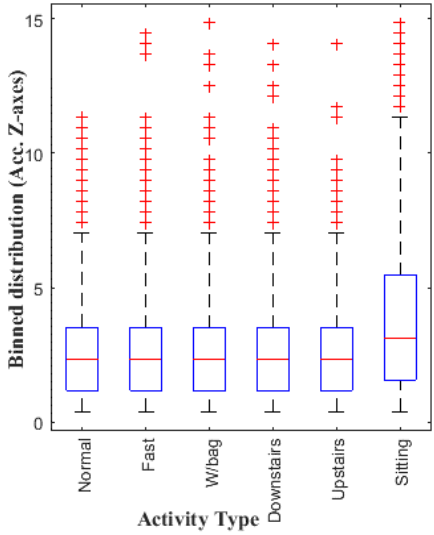

Figure 7: Binned distribution of the accelerometer z-axis feature.

When plotting the data points by transforming the top 10 ranked features using the PCA algorithm, the activity data points tend to be located close to each other in the PCA feature space (Goodall and Jolliffe, 2002; Bro and Smilde, 2014). Figure 8 shows the dataset observations for the six activities utilising the first three principal components. The first, second and third PCs used in this plot explain 72.5%, 11.7% and 10.8% of the total variance, respectively, the total variance being the sum of the variances of all the individual PCs.

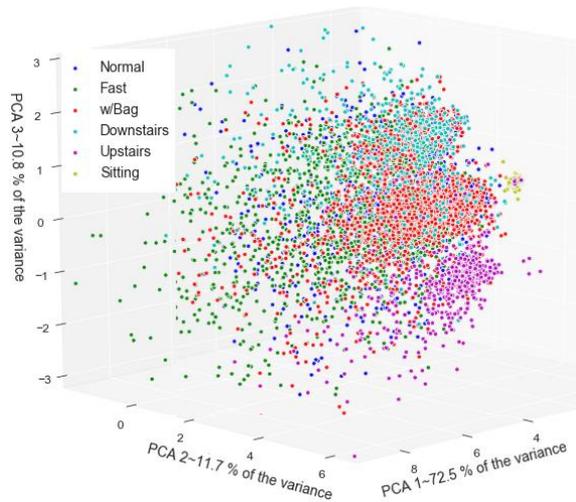

Figure 8: PCA data points scattered using the top 10 ranked features.

Table 5: Overall classification accuracy for each model.

| Activity type | Activity merge | #Features | XGB (%) | SVM (%) | NN (%) | Soft voting (%) | Hard voting (%) |
|---|---|---|---|---|---|---|---|
| Normal, Fast, Downstairs, Upstairs, Sitting | W/bag merged with Normal | 195 | 93.54 | 93.35 | 93.88 | **94.60** | 94.27 |
| | | 304 | 93.11 | 93.30 | 93.50 | 94.17 | 93.64 |
| Walk, Downstairs, Upstairs, Sitting | Fast and W/bag merged with Normal | 195 | 97.06 | 97.30 | 97.65 | **97.79** | 97.54 |
| | | 304 | 97.01 | 97.15 | 97.49 | 97.73 | 97.49 |
| All | None | 195 | 86.18 | 84.88 | 87.67 | **87.79** | 87.24 |
| | | 304 | 84.64 | 84.54 | 84.83 | 87.14 | 86.95 |

It is apparent that some outliers sit far from their group or overlap with another group, and these could be misclassified by the predictable model built. However, outliers were included in the classification tests and were not excluded from any process within this experiment, as they are real-world samples.

Three different experimental settings were undertaken to study how various activity types affected the identification rate. First, as normal walk and walk with bag are the most similar activity types, they were merged to form a single activity. The second test merged normal, fast, and walk with bag into a single activity. The final test examined the correct classification rate for all the activities. Two types of voting were used: hard and soft majority voting.

Using only an accuracy metric does not fully reveal overlapping and false positive rates among the classes, as it computes the ratio of true predicted labels to the total examined sample, which becomes insensitive to unbalanced classes. Therefore, an F score is computed, which is interpreted as the weighted mean of the precision and recall. An F score of 1.0 is the highest and a lowest score of 0.0 is the lowest. It worth mentioning that it is common to use F score for binary classification problems, however, adapting the metric for multiclass problem is achieved using one label versus all other labels. In which, the relative contribution of precision and recall to the F score is equal, as Figure 9 illustrates.

Figure 9 shows that sitting and walking upstairs have the highest F-score. In contrast, walking downstairs has the lowest recall and F-score rates in comparison with the other activities. Also, the figure shows the support of each class which represents the number of occurrences of each class in the test set.

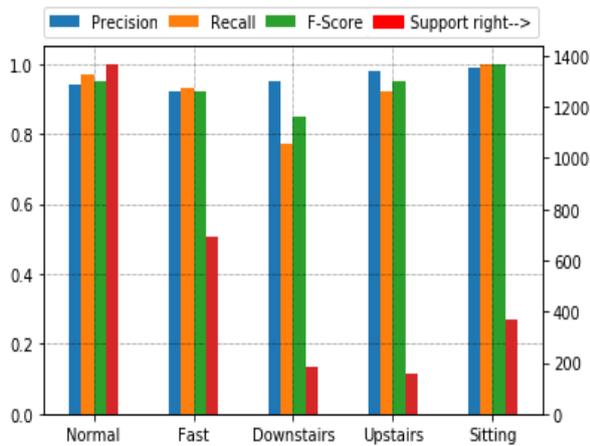

Figure 9: Precision, recall, F-score and support of five activities (walk with bag samples are merged with normal walk).

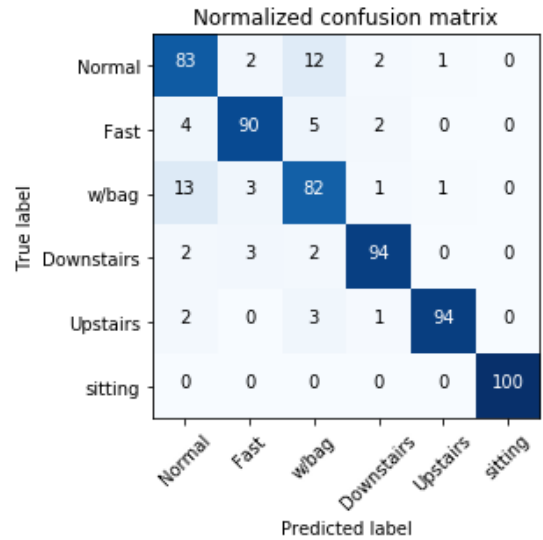

Figure 10: Normalised confusion matrix (%) of the soft voting model.

The confusion matrix summarises the performance of the classification model for the multi-class classification task in this study (in particular the soft voting model). It also shows how the predictable model performs on a class level, in which both true-positive and false-negative values can be measured. presents the normalised confusion matrix for the percentages for all six activities. It is not surprising that sitting has the highest prediction rate of the activities. This is due to the uniqueness of its generated sensor signals, as in both the top- and low-ranked features, it was clearly distinguished from the other activities.

The finding is also consistent with the box-plots in Figure 6 and Figure 7, in which the misclassified samples of the normal walk activity are mostly assigned to the walk with bag activity and vice versa. With regard to the downstairs activity, the false-positive samples are misclassified as walking types (either normal, fast or with bag) and this could be interpreted as some of the downstairs samples actually containing normal and fast walk types. For example, once a subject reaches the bottom of the stairs, the individual walks a few more steps to complete the activity, which might become a noisy/outlier sample in the downstairs activity dataset.

## 5. DISCUSSION

One of the most interesting findings of this study is that all the top 10 ranked features, as illustrated in Figure 5, are based on only the y-axis of the gyroscope and accelerometer sensors. This could be interpreted as being due to the location of the device, as it was placed on the side of the person's waist, which makes the y-axis the axis most sensitive to human walk-based activity motions. It would be interesting to assess the effects of different device locations during sensing using the same experimental setup proposed in this study. Although the developed approach reached a high level of accuracy in identifying human physical activity based on raw smartphone motion sensor signals, other aspects could be examined and investigated in future research to generate more findings, including the following:

- The evaluation of this study was conducted offline using a desktop computer. It has not been thoroughly tested in a live environment (smartphone) to measure other operational metrics, such as computational overheads, memory consumption and the time required for the whole pipeline to be completed, starting from acquiring motion signals, to feature extraction, segmentation, pre-processing, and finally inferencing, where the examined data are classified into the right activity type.
- The collected dataset was acquired using a single type of mobile device (Samsung Galaxy S6). Investigating other widely used devices, such as an Apple iPhone, could reveal how similar/different the generated motion signals might be for different devices and to what extent feature space distribution varies.
- Future work could also investigate other factors, such as identifying the minimum number of seconds and samples required per individual in order to train a user-dependent predictable model successfully in order that it can accurately match a given signal with the corresponding physical activity. This study constructed a general predictable model that takes advantage of the signals generated by the whole dataset population (60 participants).

# 6. CONCLUSIONS

The findings of this study provide evidence that it is possible to identify an individual's physical activity with a high degree of accuracy, reaching nearly 98%, based on smartphone-embedded gyroscope and accelerometer sensor signals gathered over two days. This was achieved by leveraging the capabilities of machine learning algorithms in two stages: feature ranking, in which the feature space is ranked based on the multiclass classification approach, followed by activity identification, in which only top-ranked features are included within the classification phase. The soft majority voting approach provides the highest accuracy in comparison with other models, such as single classifier or hard majority voting.